\documentclass[10pt,conference]{IEEEtran}
\IEEEoverridecommandlockouts
\usepackage{cite}
\usepackage{amsmath,amssymb,amsfonts}
\usepackage{algorithmic}
\usepackage{graphicx}
\usepackage{textcomp}
\usepackage{xcolor}
\def\BibTeX{{\rm B\kern-.05em{\sc i\kern-.025em b}\kern-.08em
    T\kern-.1667em\lower.7ex\hbox{E}\kern-.125emX}}

\usepackage{enumerate}

\usepackage{url}
\usepackage{todonotes}
\usepackage{nameref}

\newcommand{\cfix}{\ensuremath{C_{fix}}}
\newcommand{\ifix}{\ensuremath{I_{fix}}}
\newcommand{\cinduce}{\ensuremath{C_{induce}}}
\newcommand{\iinduce}{\ensuremath{I_{induce}}}

\def\mrqa{How often are simple bugs fixed by a different author?}
\def\mrqb{Does bug fix authorship affect the bug fix time?}
\def\mrqc{Does bug fix authorship affect the commit size of simple bug fixes?}

\begin{document}


\title{Mea culpa: How developers fix their own simple bugs differently from
    other developers}

\author{\IEEEauthorblockN{Wenhan Zhu and Michael W. Godfrey} 
\IEEEauthorblockA{\textit{David R. Cheriton School of Computer Science} \\
\textit{University of Waterloo}\\
Waterloo, Ontario, Canada \\
\{w65zhu,migod\}@uwaterloo.ca}
}

\maketitle

\begin{abstract}
    In this work, we study how the authorship of code affects bug-fixing
    commits using the \emph{SStuBs} dataset, a collection of
    single-statement bug fix changes in popular Java Maven projects.
    More specifically, we study the differences in characteristics between
    simple bug fixes by the original author --- that is, the developer who
    submitted the bug-inducing commit --- and by \emph{different}
    developers (i.e., non-authors).
    Our study shows that nearly half (i.e., 44.3\%) of simple bugs are
    fixed by a different developer. 
    We found that bug fixes by the original author and by different
    developers differed qualitatively and quantitatively.
    We observed that bug-fixing time by authors is much shorter than
    that of other developers.
    We also found that bug-fixing commits by authors tended to be larger in
    size and scope, and address multiple issues, whereas bug-fixing commits
    by other developers tended to be smaller and more focused on the bug
    itself.
    Future research can further study the different patterns in bug-fixing
    and create more tailored tools based on the developer's needs.
    
\end{abstract}


\section{Introduction}

Research has shown that both bugs~\cite{catolino2019not} and bug
fixes~\cite{pan2009toward} have different patterns. For example, fixing
GUI related bugs is different from fixing database related bugs. Fixing GUI
bugs often involves manual inspection of the GUI and is hard to automate.
On the other hand, database related bugs often require extra attention to
avoid damaging the integrity of the underlying database.
A deeper understanding of how and why bugs occur can help developers focus
on the weaker aspects of the system and its development practices to
produce better software in the long run~\cite{rahman2011ownership}.
Large software systems require many developers to contribute in different
subsystems.
As the software system evolves, the set of developers working on the
project also changes; over time, developers will often edit code that was
originally written by another developer.
The authorship of code can make a huge difference in software quality and
bug prediction~\cite{hu2014effective,d2010extensive,di2017developer}.

In this work, we use the \emph{SStuBs}\cite{sstubs} dataset to better
understand how bug fixes differ when they are fixed by the developer who
wrote the original code (the ``author'') compared to when they are fixed by
another developer.
While differences in bug-fixes have been studied from the code
perspective~\cite{pan2009toward}, we examine it from the developer
perspective.
Moreover, previous studies on the characteristics of bug fixes have mostly
based on a small sample size from a few projects \cite{catolino2019not} with
a few hundred samples \cite{wen2019exploring}. 
Our work here uses the \emph{SStuBs} dataset, which comprises of 10,231
instances of singe-statement bug fixes across the top 100 Java Maven
projects. 
This allows us to investigate bug fixes at a larger scale both in the
number of projects and in the number of samples compared to previous
studies.

In this paper, we perform an empirical study on the differences in
bug-fixing commits for simple bugs across two dimensions:  bug fixes
submitted by the original author of the code, and bug fixes submitted by
other developers.  We compare the bug fixes in terms of size and scope of
the commit, and the time taken to fix.
We address three research questions:

\begin{enumerate}[\textbf{RQ}1]
    \item \textbf{\mrqa} \\
    We find developers fix simple bugs from another developer's code in
    44.3\% of the cases.
    \item \textbf{\mrqb} \\
    Developers fix simple bugs in their code faster --- with a median of
    time of less than one day --- compared to fixing simple bugs from
    another developer, with a median time of 148 days.
    \item \textbf{\mrqc} \\
    Simple bug fixes by the same developer have larger commit size with
    a wide variation in range: we found an interquartile range (IQR) of 734
    LOC from 4 LOC at the first quantile to 738 LOC at the third quantile.
	Meanwhile, simple bug fixes by a different developer are small and
	vary less: we found an IQR of 13 LOC from 2 LOC at the first
	quantile to 15 LOC at the third quantile.
\end{enumerate}

Our study suggests that bug fixes by different developers exhibit different
patterns:
Bug fixes by the same developer tend to occur within a short amount of time
of the original bug-inducing commit, and are usually embedded within
a larger commit.
By contrast, bug fixes by a different developer tend to happen later in
time, and the commit that fixes the bug tends to be confined in scope to
the bug fix itself.

\section{Data Collection}

In this work, we use the \emph{SStuBs}\cite{sstubs} dataset which contains
simple-statement bug fixes. We use the variation that contains 10,231 bug
fixes from the top 100 Java Maven projects in the dataset as the basis for
our analysis.  We also provide a full replication package for our
work\footnote{\url{https://anonymous.4open.science/r/344cf208-ea32-49f4-90fe-59bdb6e5d7fe/}}.

We cloned every project from the list of top 100 Maven projects from the
\emph{SStuBs} dataset in Jan 2021. One of the projects \emph{b3log/solo}
has moved to another location to \emph{88250/solo}, so we opted to use the
later repository. The change of location does not affect the history of the
commit, so it does not affect our analysis.

\section{Methodology}

\begin{figure}[htbp]
    \centerline{\includegraphics[width=0.5\textwidth]{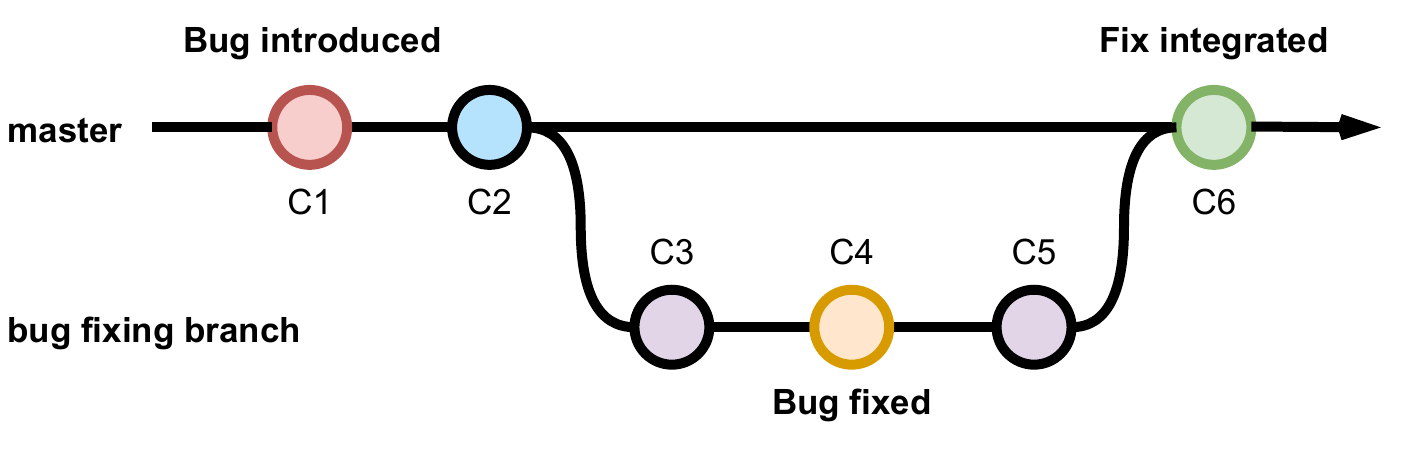}}
    \caption{Illustration of separate fix and integration commit}
\label{fig:git_flow}
\end{figure}

\subsection{Bug-inducing and bug-fixing commits}
\emph{SStuBs} already stores the bug-fixing commit and the last commit
where the bug still exists in the codebase using the SZZ algorithm
\cite{sliwerski2005changes}.
We go one step further and use a slightly modified version of the SZZ
algorithm to locate the authorship of the bug-inducing and bug-fixing
commits by ignoring merging commits. 
As shown in Fig.~\ref{fig:git_flow}, when applying the SZZ algorithm,
\emph{C6} will be identified as the bug-fixing commit.
However, \emph{C6} is the branch merging commit, therefore does not
represent the actual commit that implements the bug fix. In this case, the
bug fix is implemented in another commit, \emph{C4}.
In practice, this typically happens when a developer implements the bug fix
in another branch and they submit a pull request to the codebase. When the
pull request is accepted, a merge commit will be created which adds the
bug-fixing change into the codebase. 
Since we are interested in the authorship of both the bug-inducing and
bug-fixing commit, we make the distinction between the implementation and
integration of the bug-fixing and bug-inducing change.
In the rest of the paper, we use the following abbreviations to refer to
the bug-fixing related commits.
Note that \cfix{} and \ifix{} can refer to the same commit. This is also
true for \cinduce{} and \iinduce{}.

\begin{itemize}
    \small
    \item \cfix{}: Bug fix integration commit
    \item \ifix{}: Bug fix implementation commit
    \item \cinduce{}: Bug induce integration commit
    \item \iinduce{}: Bug induce implementation commit
\end{itemize}

We removed 11 bug fixes where their related commits do not exist anymore in
the corresponding \emph{git} repositories. The missing commits can be
caused by deleted branches in the repository or lost from pull requests
from another no longer available source\footnote{It is a common practice to
delete no longer required forks if the changes integrate back.}. 

\subsection{Author determination}
With both the integration and implementation commits available for
bug-inducing and bug-fixing changes, the authorship of both changes can be
easily determined by investigating the commit information.
In this work, we are mostly interested in the developers writing the
bug-inducing and bug-fixing changes, so we use the authorship information
from both the implementation commits, \ifix{} and \iinduce{}.

\subsection{Bug fix time}
The bug fix time refers to the time between the bug-inducing commit and the
bug-fixing commit. Unlike the authorship of the code change where we need to
trace to the implementation commit, bug fixes relative to the project
development timeline. Therefore, we use the time difference between \cfix{}
and \cinduce{} as a measurement for the bug fix time. Specifically, we use
$commit\_time(\cfix{}) - commit\_time(\cinduce{})$ to calculate the bug fix
time for each bug.

\subsection{Code Churn}
In this paper, code churn is considered as the number of lines of code
(LOC) changed in a commit. Since simple bugs in \emph{SStuBs} refer to bugs
that occur within a single statement, a bug-fixing change often modifies
only a single line and thus not change the total number of lines in the
codebase.  Consequently, when calculating the code churn as $lines\_added
- lines\_removed$, a large proportion of bug fixes will have a net code
churn of zero. In fact, 68.5\% (i.e., 7,013) of all simple bugs in the
dataset exhibit this property.
Hence, to capture more information in the change, we use an alternative
calculation of code churn as $abs(lines\_added) + abs(lines\_removed)$ to
reserve information on the total lines modified.

\section{Results}
\subsection{RQ1: \mrqa}

In industry, developers often need to fix bugs written by others.  This is
particularly evident in open source systems, where developers are often
volunteers whose commitment and participation levels may vary over time.
In this RQ, we explore how often simple bugs are fixed by a developer other
than the original author.

For each bug-fixing commit, by comparing \ifix{} and \iinduce{}, we can
determine whether the bug is fixed by the same developer who contributed
the original code.
Using our modified version of the SZZ algorithm in tracing the
implementation commit, we find that 44.3\% (i.e., 4,508) of simple bug
fixes are from developers fixing bugs in another developer's code.
This observation shows a non-trivial amount of bug fixes by a different
developer contributing to the understanding of developer bug-fixing activities.
Researchers should be aware of the difference in developer fixing the bug
as they can represent different workflow and therefore require different
attention.

\begin{table}[htbp]
    \caption{Bug fixes by author}
\begin{center}
\begin{tabular}{|c|c|c|}
\hline
Total bug fixes & Fixed by same author & Fixed by different author \\
\hline
10,182 & 5,674                & 4,508                     \\
\hline
\end{tabular}
\label{tab:bug_fix_by_author}
\end{center}
\end{table}

\subsection{RQ2: \mrqb}
Despite the hope of producing perfect software with no bugs, during
development in practice, bugs can not be prevented. 
Alternatively, there has been a substantial effort to improve the bug-fixing
quality and time. High-quality bug reports~\cite{bettenburg2008makes} for
example are a useful asset for developers when fixing bugs.
In this RQ, we study how the difference in the authorship of the bug fix
affects the bug fix time. 
As discussed above, we consider the bug fix time from the project
perspective. 
More specifically, we consider the bug fix time as the time from when the
buggy code is committed to the code base to the time the bug fix change is
committed to the code base.

As shown in Fig.~\ref{fig:bug_fix_time_by_author}, simple bug fixes by
different authors (with a median of 148 days) occur much longer compared to
simple bug fixes by the same author (with a median of less than 1 day).
The large difference in bug fix time suggests these may be inherently
different patterns of development activity. 
For example, the short fix time of simple bugs for the same author might be
an artifact from the normal development process. How often and how much to
commit is an on-going argument in the development process. For example,
even with continuous integration tools (CI), it is hard to ensure every
commit of a project resembles a running state. The enforcement of a
complete project is often only ensured at releases or pull requests. In
this case, the simple bug fixes by the same author may be a result of the
artifact of the normal development process.
On the other hand, simple bug fixes by a different author  have a
wider time range with a median bug fix time of 148 days.
In future work, a qualitative study can be performed to better understand
the intent and cause of simple bugs to explain the large difference in the
bug fix time between the same and a different author.

\begin{figure}[htbp]
    \centerline{\includegraphics[width=0.5\textwidth]{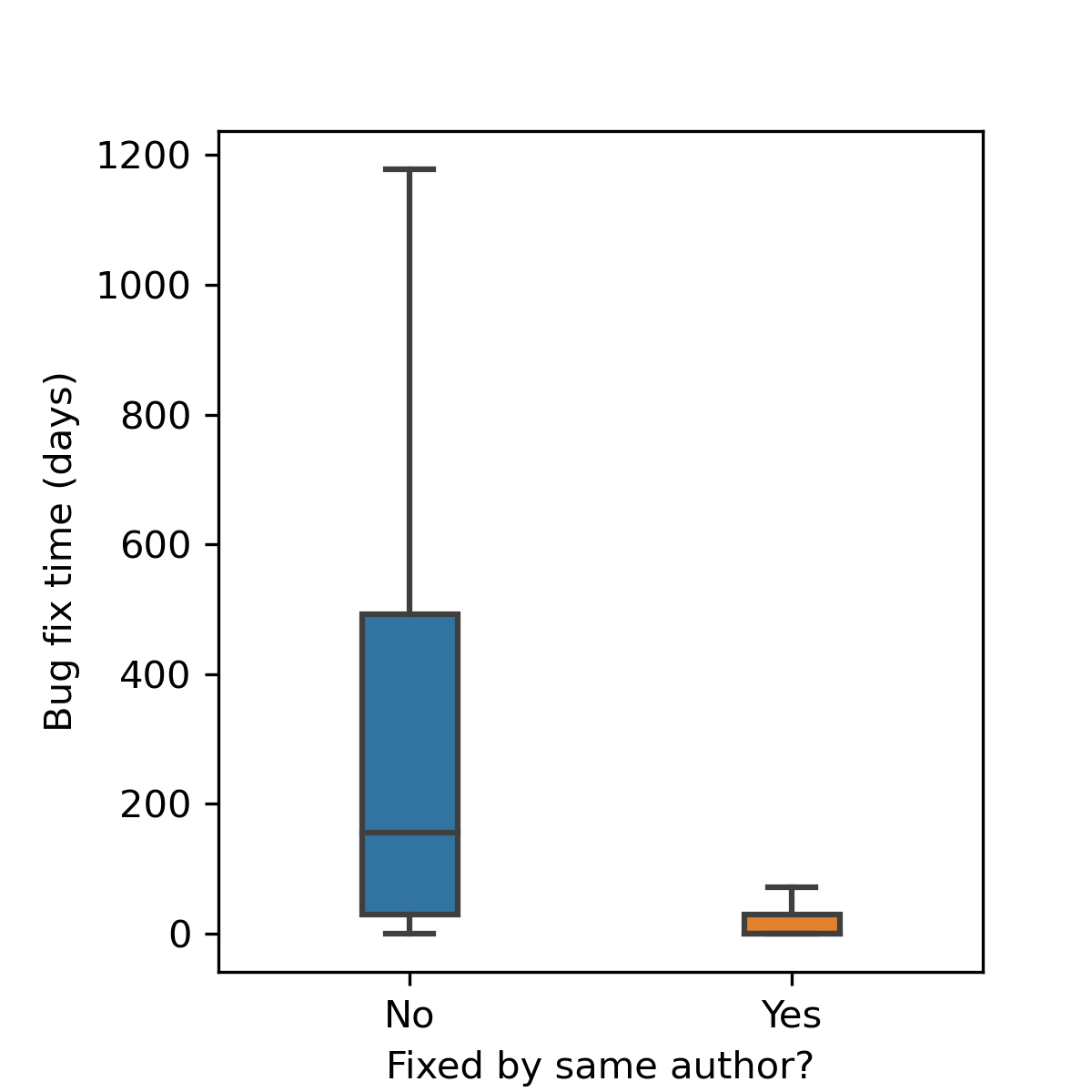}}
\caption{Bug fix time by same author and different author}
\label{fig:bug_fix_time_by_author}
\end{figure}

\subsection{RQ3: \mrqc}

Following RQ2 where we discovered a difference in bug fix time between
simple bug fixes by the same and a different author, we continue to
investigate whether there are also differences in bug fix size.
In theory, each commit should do one thing. However, this is often not the
case in practice. Developers often combine multiple things in one commit
and do not follow best practices~\cite{bird2009promises}. When these
situations happen, it is often hard to untangle the commit.
In this RQ, we explore the bug fix size of simple bug fixes to explore
whether there are differences between bug fixes by the same developer and a
different developer.

Fig.~\ref{fig:bug_fix_churn_by_author} shows the churn of simple bug fixes
by the same author and a different author. Simple bug fixes by
different authors are typically small (e.g., a median of 6 LOC) compared to
a larger range of commit size of simple bug fixes by the same author (with
a median of 14 LOC). 
Bug-fixing commits by same developers also vary by a large amount with the
interquartile range (IQR) of 734 LOC ranging from 4 LOC at the first
quantile to 738 LOC at the third quantile.
Meanwhile, simple bug fixes are smaller and have a small range by a
different developer. The IQR is only 13 LOC ranging from 2 LOC to 15 LOC
from the first to third quantile.
The observed difference in commit size echoes our finding in RQ2 that simple
bug fixes by the same developer may represent a different pattern than the bug
fixes by a different developer.
The large size in the bug-fixing commit suggests that simple bug fixes are
embedded in a larger commit and therefore the simple bug fix may not be the
main purpose of the commit. On the other hand, the bug-fixing commits by a
different developer is relatively small in size, suggesting the purpose of
these bug-fixing commits are more pin-pointed at fixing the simple bugs.

\begin{figure}[htbp]
    \centerline{\includegraphics[width=0.5\textwidth]{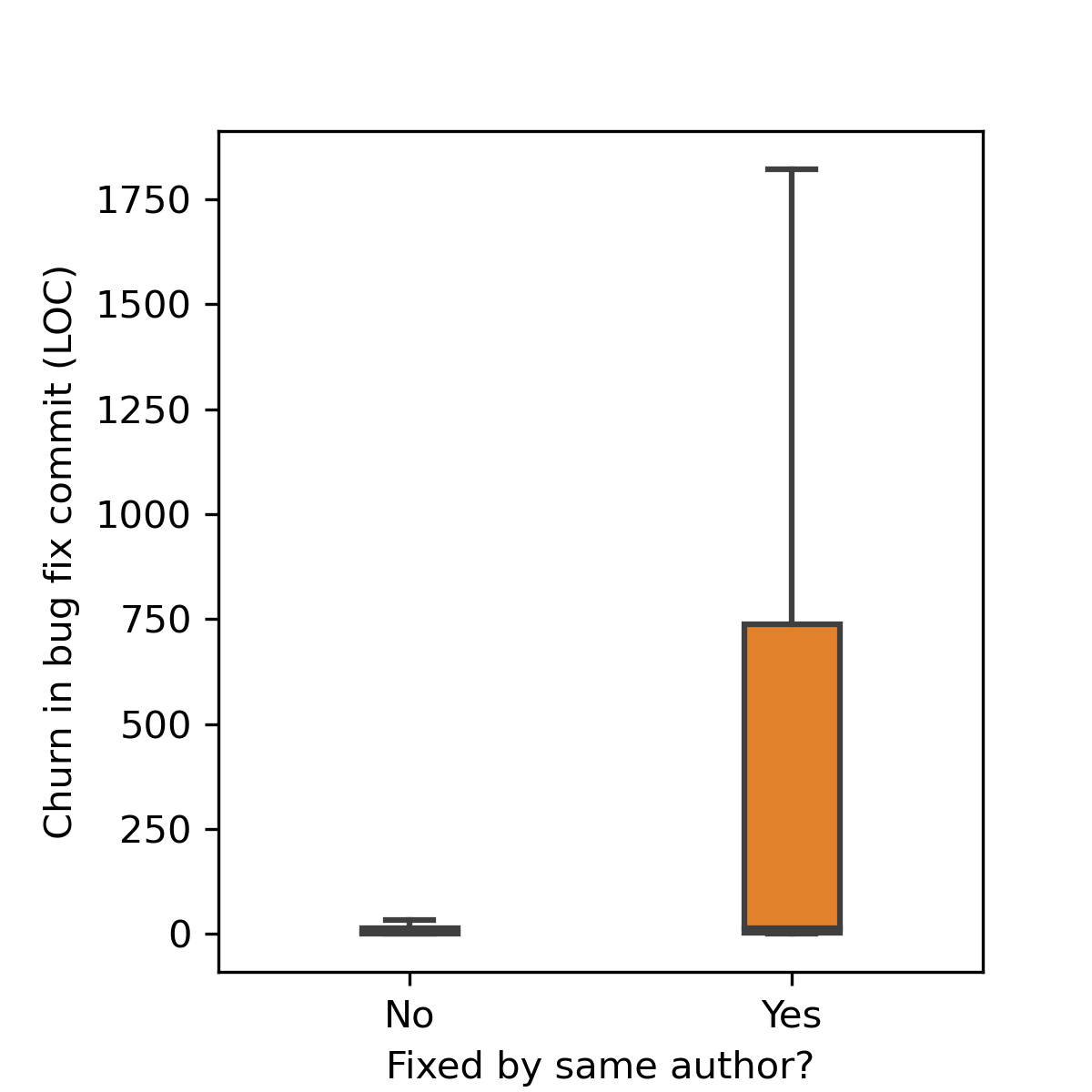}}
    \caption{Code churn in bug-fixing commits by authorship}
    \label{fig:bug_fix_churn_by_author}
\end{figure}

\section{Discussion}
Our observations show that simple bug fixes have different patterns
depending on whether the bug is fixed by the same developer writing the
original code.
When developers are fixing simple bugs from their code, they tend to
fix the bug quickly in a short amount of time and often include the bug fix
in a larger patch. On the other hand, bug fixes by a different developer
tends to be small and often occur a long time after the original code is
introduced.
The difference in characteristics suggests that simple bug fixes by the
same author may be inherently different from bug fixes by a different
developer.  We raise the following theory on why this occurs.
Our observation suggests that simple bugs are fixed by the same developer
quickly with a median time of less than one day. During development, the
developer may not be able to perform large scale testing at every stage to
ensure the contributed code functions correctly. After the rapid
development phase with many changes in source code with commits recording
the changes, simple bugs will appear and be fixed promptly. And when the
period of development is finished, some simple bugs will be caught and
fixed quickly. Therefore, resulting in our observation of quick simple bug
fixes by the same developer.  On the other hand, bug fixes by a different
developer do not go through the development cycle and therefore do not
suffer from the artifacts resulting in a more natural distribution of bug
fix times. Our observation on the larger commits containing simple bug
fixes by the same author also supports this theory.

As the bug-fixing activities are different for authors and non-authors,
future research can further investigate the intent and cause of simple
bugs. With a better understanding of why they happen, researchers can build
more intelligent tools to help developers.
For example, when predicting bugs, tools may consider the factor on whether
the piece of code have been modified by another developer as suggested by
our work that such activity is different than modifying code by the same
developer.
Another example, as one of the goals for the \emph{SStuBs} dataset is to
evaluate program repair techniques, our observations indicate that not
every simple bug fix in the dataset is a good target for program repair
tools as the bug fixes by the same author may come from artifacts during
software development and therefore do not necessarily represent a good
bug-fixing example.

\section{Threats to Validity}
\subsection{Internal Validity}

\subsubsection{Determine Implementation Commit}
We rely on the \emph{git-blame} build-in command and the bug fix location
information from \emph{SStuBs} to retrieve the history of source code
modification. However, the history of \emph{git} can be overwritten and the
bug fix location may not be precise from \emph{SStuBs}. Therefore yielding
inaccurate implementation commit being determined affecting our accuracy in
bug fix time and code churn measured. 
Moreover, the bug may have been induced unrelated to the bug fix location,
hence not correctly representing the actual bug-fixing process.

\subsubsection{Determining Commit Authorship}
We used the author's e-mail as the unique identifier for the author's
identity. However, in practice, the same developer may use different
configuration files on different machines, resulting in two different
identifier mapping to the same developer. Consequently, during our
analysis, we may have bug fixes by the same author categorized as different
authors.

\subsection{External Validity}
Our study only looked at simple bug fixes in top \emph{Maven} projects in
\emph{Java} managed through \emph{git}, therefore may not be generalised to
other projects using different languages or different version control
systems.
Our study also may not be generalized to other types of bugs as study has
shown not all bugs are the same~\cite{catolino2019not}.

\section{Conclusion}
We conducted an empirical study on how the authorship affects simple bug
fixes from the \emph{SStuBs} dataset. We traced the bug fixes to determine
the developers writing and fixing the buggy code. We observe that
developers fix simple bugs from another developer's code in 44.3\% of the
cases. 
Our result shows that when developers are fixing simple bugs in their own
code, they tend to fix quickly and often embed the bug fix in a larger
commit.
On the other hand, simple bug fixes by a different developer tend to occur
later in time, and the bug fix commit tends to be confined in scope to the
bug fix itself.
Our observations indicate different patterns in fixing simple bugs from
the authorship. Future work on bug-fixing should consider incorporating this
information when designing tools that better suit the developer's needs.

\bibliographystyle{IEEEtran}
\bibliography{references}

\begin{thebibliography}{10}
\providecommand{\url}[1]{#1}
\csname url@samestyle\endcsname
\providecommand{\newblock}{\relax}
\providecommand{\bibinfo}[2]{#2}
\providecommand{\BIBentrySTDinterwordspacing}{\spaceskip=0pt\relax}
\providecommand{\BIBentryALTinterwordstretchfactor}{4}
\providecommand{\BIBentryALTinterwordspacing}{\spaceskip=\fontdimen2\font plus
\BIBentryALTinterwordstretchfactor\fontdimen3\font minus
  \fontdimen4\font\relax}
\providecommand{\BIBforeignlanguage}[2]{{%
\expandafter\ifx\csname l@#1\endcsname\relax
\typeout{** WARNING: IEEEtran.bst: No hyphenation pattern has been}%
\typeout{** loaded for the language `#1'. Using the pattern for}%
\typeout{** the default language instead.}%
\else
\language=\csname l@#1\endcsname
\fi
#2}}
\providecommand{\BIBdecl}{\relax}
\BIBdecl

\bibitem{catolino2019not}
G.~Catolino, F.~Palomba, A.~Zaidman, and F.~Ferrucci, ``Not all bugs are the
  same: Understanding, characterizing, and classifying bug types,''
  \emph{Journal of Systems and Software}, vol. 152, pp. 165--181, 2019.

\bibitem{pan2009toward}
K.~Pan, S.~Kim, and E.~J. Whitehead, ``Toward an understanding of bug fix
  patterns,'' \emph{Empirical Software Engineering}, vol.~14, no.~3, pp.
  286--315, 2009.

\bibitem{rahman2011ownership}
F.~Rahman and P.~Devanbu, ``Ownership, experience and defects: a fine-grained
  study of authorship,'' in \emph{Proceedings of the 33rd International
  Conference on Software Engineering}, 2011, pp. 491--500.

\bibitem{hu2014effective}
H.~Hu, H.~Zhang, J.~Xuan, and W.~Sun, ``Effective bug triage based on
  historical bug-fix information,'' in \emph{2014 IEEE 25th International
  Symposium on Software Reliability Engineering}.\hskip 1em plus 0.5em minus
  0.4em\relax IEEE, 2014, pp. 122--132.

\bibitem{d2010extensive}
M.~D'Ambros, M.~Lanza, and R.~Robbes, ``An extensive comparison of bug
  prediction approaches,'' in \emph{2010 7th IEEE Working Conference on Mining
  Software Repositories (MSR 2010)}.\hskip 1em plus 0.5em minus 0.4em\relax
  IEEE, 2010, pp. 31--41.

\bibitem{di2017developer}
D.~Di~Nucci, F.~Palomba, G.~De~Rosa, G.~Bavota, R.~Oliveto, and A.~De~Lucia,
  ``A developer centered bug prediction model,'' \emph{IEEE Transactions on
  Software Engineering}, vol.~44, no.~1, pp. 5--24, 2017.

\bibitem{sstubs}
R.-M. Karampatsis and C.~Sutton, ``How often do single-statement bugs occur?
  the manysstubs4j dataset,'' in \emph{Proceedings of the International
  Conference on Mining Software Repositories (MSR 2020)}, 2020.

\bibitem{wen2019exploring}
M.~Wen, R.~Wu, Y.~Liu, Y.~Tian, X.~Xie, S.-C. Cheung, and Z.~Su, ``Exploring
  and exploiting the correlations between bug-inducing and bug-fixing
  commits,'' in \emph{Proceedings of the 2019 27th ACM Joint Meeting on
  European Software Engineering Conference and Symposium on the Foundations of
  Software Engineering}, 2019, pp. 326--337.

\bibitem{sliwerski2005changes}
J.~{\'S}liwerski, T.~Zimmermann, and A.~Zeller, ``When do changes induce
  fixes?'' \emph{ACM sigsoft software engineering notes}, vol.~30, no.~4, pp.
  1--5, 2005.

\bibitem{bettenburg2008makes}
N.~Bettenburg, S.~Just, A.~Schr{\"o}ter, C.~Weiss, R.~Premraj, and
  T.~Zimmermann, ``What makes a good bug report?'' in \emph{Proceedings of the
  16th ACM SIGSOFT International Symposium on Foundations of software
  engineering}, 2008, pp. 308--318.

\bibitem{bird2009promises}
C.~Bird, P.~C. Rigby, E.~T. Barr, D.~J. Hamilton, D.~M. German, and P.~Devanbu,
  ``The promises and perils of mining git,'' in \emph{2009 6th IEEE
  International Working Conference on Mining Software Repositories}.\hskip 1em
  plus 0.5em minus 0.4em\relax IEEE, 2009, pp. 1--10.

\end{thebibliography}

\end{document}
